# The Chirality Trinity: Chirality, Chirality Prime and Time Chirality

Sang-Wook Cheong[1,*] and Fei-Ting Huang[1]

[1]Keck Center for Quantum Magnetism and Department of Physics and Astronomy, Rutgers University, Piscataway, New Jersey, 08854, USA

**ABSTRACT**
The flow of time moves in one direction in any spatial position and orientation in this universe. Chiral objects, which lack mirror symmetry, retain their chirality regardless of their position or orientation. Despite being seemingly independent, time and chirality share common features such as universality—applying to "any position and orientation"—and a binary nature, such as forward/backward time flow versus left/right chirality. We introduce the concept of Time Chirality and discuss the conjugate relationship between time chirality and traditional chirality. We explore how time chirality can manifest in certain magnetic states, and examine the novel physical phenomena associated with time-chiral magnetic states. This discussion offers a fresh perspective on true time reversal symmetry breaking and temporal nonreciprocity.

## I. INTRODUCTION

Our universe exhibits Temporal nonreciprocity, in the sense that, for example, we everyday experience that time cannot be reversed, i.e., the flow of time moves only in one direction, and the disorder or entropy of an isolated system will always increase or remain constant over time[1-4]. On the other hand, the fundamental physical processes governing equilibrium states, with no net fluxes of quantities such as energy, particles, or momentum, often remain unchanged if the direction of time is reversed. For example, the overall distribution and energy density in thermal equilibrium blackbody radiation will remain unchanged when time is reversed. Another example is: a perfectly elastic material under stress will deform but return to its original shape when the stress is removed. In time reversal, the process of deformation and return to shape will look the same. In fact, the basic equations of motion in both classical and quantum mechanics, such as Newton's laws, Maxwell's equations, and the Schrödinger equation, are time-reversible[5-10].

Chirality[11-17] describes a property where an object or a state and its mirror image cannot be superimposed, regardless of spatial rotations, translations, and even time-reversal symmetry (**T**-symmetry). This means that the object lacks mirror symmetry in any spatial orientation or position. Space inversion can be identical with a combination of a mirror reflection and the relevant 2-fold spatial rotation, this reinforces the absence of space inversion (parity (**P**)). Thus, chirality lacks **P** as well as **PT** (**P** times **T**) symmetries. In the case of magnetic chirality[18], a similar principle applies: no symmetry operations other than mirror reflection and space inversion can superpose chiral magnetic objects/states with their mirror images[14, 15]. Louis Pasteur is often considered the father of molecular chirality, or molecular dissymmetry. His pioneering work laid the foundation for the concept. Later, Lord Kelvin coined the term 'chirality'[11-15] further advancing the understanding of this fundamental property. Chiral functionalities encompass natural optical activity, magnetochiral effect[19, 20], longitudinal current-induced magnetization[21], chirality-selective spin-polarized current of charged electrons or neutral neutrons[22], chiral phonons[23, 24], nonlinear optical effect[25, 26], chiral-selective (surface) chemical reactions[27, 28], and shear-type piezoelectricity[29] - many of which are critical for numerous technological applications. In this perspective, we introduce the term Time chirality and discuss the conjugate and common nature between the time reversal and the mirror reflection of chirality, and emergent properties and phenomena from this unique consideration. The common nature includes the binary character of forward or backward time flow vs. left or right chirality.

## II. METHODS

Symmetry operation similarities (SOS)[30]: The relationships between specimen constituents (specimens (lattice distortions, spin arrangements, etc.) under certain experimental setups such as applied electric/magnetic fields, light illumination, etc.) and observables (measuring quantities such as strain, electric polarization, magnetization, etc.) are analyzed in terms of the characteristics under various symmetry operations of rotation, space inversion, mirror reflection, and time reversal. When specimen constituents and observable share the same broken symmetries, except translation symmetry, they are said to exhibit symmetry operation similarity (SOS), and the corresponding phenomena can occur. "A has SOS with B" means that A has the same symmetry with or a lower symmetry than B, *i.e.*, A does not have more unbroken symmetries than B does. Emphasize that this SOS approach is a practical mean to utilize Neumann's principle[31] in real experimental situations.

*Contact author: sangc@physics.rutgers.edu

Free rotation condition: We consider symmetry operations in combination with "translations", "proper spatial rotations", or "any combinations of them", referred to as the free rotation condition. In the free rotation condition, broken **T**-symmetry means that both anti-translation (**T** combined with any translation) and anti-rotation (**T** combined with any proper spatial rotation) are broken. For example, the magnetic point group (MPG) of **4′** has the $C_{4z}$**T** symmetry, but broken **T** symmetry; however, in the free rotation condition, **4′** does have unbroken **T**-symmetry. We expand this concept of the free rotation condition to the free spatial operation condition: We consider symmetry operations in combination with "translations", "proper spatial rotations", "inversion" "mirror reflection" or "any combinations of them", referred to as the free spatial operation condition.

## III. RESULTS

### A. Time chirality

Time chirality (or temporal chirality) is defined for a property of an equilibrium state when the equilibrium state and its time-reversed state (**T**-state) cannot be identical through any spatial symmetry operations, such as translations, rotation, mirror reflection, or space inversion. This criterion is called as the free spatial operation condition (see Method). These time-chiral states, which are equilibrium states with no net fluxes, can occur through, for example, magnetic texture, artificial metamaterials, complex molecule systems or magnetic phase transitions in crystalline materials[32]. Simple ferromagnets (↑↑) or antiferromagnets (↑↓) are not time-chiral, as **T**-symmetry operation combined with certain spatial rotations and/or translation can restore the original states. The magneto-toroidal state generated by a four head-to-tail configuration of spins and a magnetic monopole with four outward spins do break **T**-symmetry, however, they remain time achiral states since **T**-states can still be reverted to the original state under the free spatial operation condition. Before introducing more complex magnetic states, we first provide simple models that explain the chiral and time-chiral states, as shown in Figure 1.

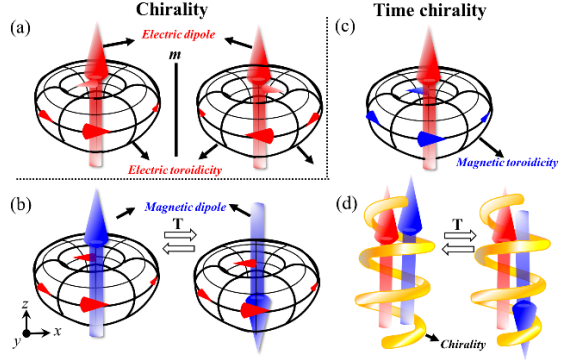

FIG. 1. Models of chirality and time chirality. (a) A chiral state characterized by an electric dipole (denoted as a vertical red arrow) and an electric toroidal moment, generated by a vortex of electric dipoles (depicted with rotating red arrows). Right panel shows its mirror state. The two states are not convertible through any spatial rotation. (b) A time-chiral state characterized by a magnetic dipole (represented by a vertical blue arrow) and an electric toroidal moment. The right panel shows its **T**-state, with two states being linked only through **T**-symmetry. (c) A time-chiral state of an electric dipole and a magnetic toroidal moment generated by a rotating spin arrangement (indicated by rotating blue arrows). (d) A time-chiral state represented by a golden chiral spring and a pair of magnetic dipole and electric dipole (left). The right panel shows its **T**-state.

First, Fig. 1(a) illustrates a chiral state created by an engineered superposition of an electric toroidal moment (depicted by rotating red arrows) and one vertical electric dipole (red arrow). This configuration has chirality, meaning it cannot be transformed into its mirror image by any spatial rotation. While only one mirror image is shown, this property holds for any mirror operation. In its mirror image (right panel), the sense of the rotation of electric dipole vortex is reversed, but the vertical electric dipole remains intact, making two states distinct and non-overlapping in the free rotation condition. Both the electric toroidal moment and the electric dipole are **T**-invariant, so this chiral state is **T**-invariant. Figs. 1(b) and 1(c) display time-chiral states, which are achieved by incorporating a **T**-variant object into the configuration of Fig. 1(a). In Fig. 1(b), a vertical magnetic dipole (blue arrow) replaces the vertical electric dipole from Fig. 1(a). The **T**-state, where the magnetic dipole is reversed but the electric toroidal moment remains, cannot be overlapped by any spatial rotation (in fact, by any spatial symmetry operations). This represents a time-chiral state distinct in the sense that it cannot be reversed or mirrored in space without violating **T**-symmetry.

In Fig. 1(c), a rotating spin arrangement (blue arrows) creates a magnetic toroidal moment, replacing the electric dipole vortex from Fig. 1(a), while the vertical electric dipole remains. This creates another time-chiral state. Finally, Fig. 1(d) illustrates additional time-chiral state created by an engineered superposition of a parallel magnetic dipole (blue arrow) and an electric dipole (red arrow) within a chiral environment. The golden chiral spring and the electric dipole are **T**-invariant, ensuring that the combined **T**-states, shown on the right, cannot be converted into the original state through any spatial symmetry operations, i.e. in the free spatial operation condition.

## B. Physical properties and phenomena of time chirality

To explore the characteristic properties and phenomena of time chirality, we utilize the concept of one-dimensional (1D) objects with their well-defined 1D directions and their dot products. The practical use of eight 1D objects is demonstrated through their dot product, as discussed in the literature [30] and detailed in the Supplemental Materials S1. It has been well established that there exist four directors: $\mathcal{D}$ (Director), $\mathcal{D}'$ (Director Prime), $\mathcal{C}$ (Chirality), $\mathcal{C}'$ (Chirality Prime), and four vectors: $\mathcal{A}$ (Ferro-rotation), $\mathcal{A}'$ (Magnetization ($M$)), $\mathcal{P}$ (Polarization), and $\mathcal{P}'$ (Velocity/linear momentum ($k$) or Magnetic Toroidal Moment)). Their dot products form the algebra of $Z_2 \times Z_2 \times Z_2$[30, 33] in terms of symmetry. Unlike director-like objects such as $\mathcal{D}'$, $\mathcal{C}$, and $\mathcal{C}'$, vector-like objects $\mathcal{A}$, $\mathcal{A}'$, $\mathcal{P}$, and $\mathcal{P}'$ point to a specific direction. Note that ['] means broken **T**. In practice, $\mathcal{C}$, $\mathcal{C}'$, $\mathcal{A}$, $\mathcal{A}'$, $\mathcal{P}$, and $\mathcal{P}'$ are, for example, associated with phenomena such as natural optical activity, linear magnetoelectricity, linear gyration, non-zero net magnetic moment, non-zero electric polarization, and nonreciprocal directional dichroism, respectively.

For $\mathcal{D}'$, we have three dot products: $\mathcal{D}' \bullet \mathcal{P} = \mathcal{P}'$, $\mathcal{D}' \bullet \mathcal{A} = \mathcal{A}'$, and $\mathcal{D}' \bullet \mathcal{C} = \mathcal{C}'$. More details are shown in Supplemental Materials S1. Each of which has significant physical meanings as illustrated in Fig. 2. First, we discuss $\mathcal{D}' \bullet \mathcal{P} = \mathcal{P}'$. On the left side of '=' symbol, the experimental setup is described, while the right side shows the observables. $\mathcal{D}'$ represents any time-chiral state. $\mathcal{P}$ refers to the application of external electric fields ($E$) or thermal gradient, while $\mathcal{P}'$ correspond to the wave vector $k$ (linear momentum, velocity) and magnetic toroidal moment. The measurable phenomena associated with $\mathcal{P}'$ include the longitudinal nonreciprocal directional dichroism (NDD) effect and transverse linear magnetoelectricity. In transverse linear magnetoelectricity, applying $E$ induces net magnetic moments that are perpendicular to the direction of $E$, and vice versa.

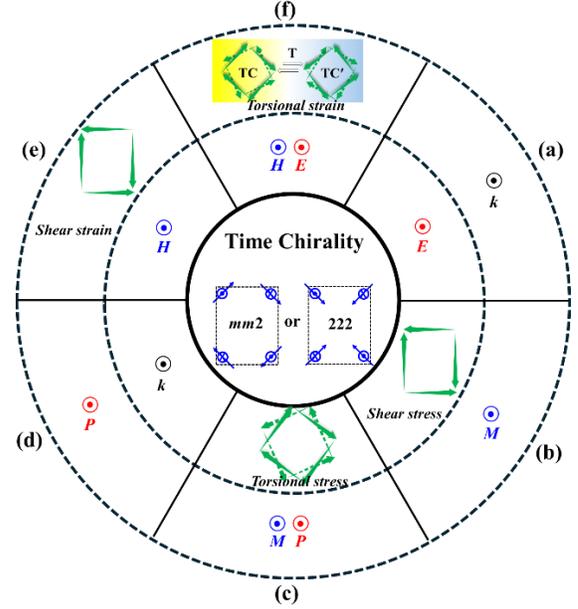

FIG 2. Various characteristic phenomena (outer circle) of time-chiral states (core) in response to external perturbations (middle circle). Two examples of time-chiral MPGs, **mm2** and **222** are shown. (a) Applying $E$ (electric field) or thermal gradient can exhibit phenomena with $k$ nature, such as Nonreciprocal Directional Dichroism and transverse linear magnetoelectricity. (b) In-plane shear force (green arrows) applied to the time-chiral state can induce magnetization $M$ along the out-of-plane axis, which is also the axis of the shear deformation. (c) Torsional stress applied to the time-chiral state can induce a linear magnetoelectricity along any direction. The top green arrows and bottom dashed green arrows rotate in opposite directions, causing a twist around the out-of-plane axis. The reverse of these effects can also occur, as shown by the opposite positions in the circle. (d) Light illumination ($k$) on the time-chiral state can induce a voltage difference ($P$) along the light propagation direction. (e) Applying magnetic field ($H$) on the time-chiral state can induce shear strain along the direction of $H$. (f) Applying both $E$ and $H$ to the time-chiral state along any direction can induce torsional strain, leading to chiral crystallographic distortions. The yellow and light blue regions represent time-chiral domains, TC and TC′, showing the opposite twisting responses.

$\mathcal{D}' \bullet \mathcal{P} = \mathcal{P}'$ indicates that applying external electric fields ($E$) or thermal gradient to any time-chiral state can induce a longitudinal NDD effect[34, 35] along the applied direction as shown in Fig. 2(a). Indeed, as predicted in Ref. 35, a real compound $Co_2SiO_4$ with time-chiral MPG *mmm* was demonstrated to exhibit an electric-field-induced NDD effect with light[36]. A more detailed discussion of the time-chiral MPG will be provided in Section C and D. Additionally, external $E$ on any time-chiral state can also induce transverse linear magnetoelectricity. We also anticipate surface transverse magnetoelectricity due to effective surface electric fields on time-chiral state surfaces.

Another powerful aspect of dot products is that they are permutable. For example, $\mathcal{D}' \bullet \mathcal{P} = \mathcal{P}'$ can be rewritten to $\mathcal{D}' \bullet \mathcal{P}' = \mathcal{P}$, implying that applying unpolarized light illumination ($k$) as $\mathcal{P}'$ on time-chiral states can induce a voltage gradient or electric polarization ($P$) along the light propagation direction (see Fig. 2(d)). We anticipate that if the incoming light consists of femto-second pulses, a photovoltaic effect or THz light emission can be induced in time-chiral states.

In fact, $\mathcal{D}' \bullet \mathcal{P}' = \mathcal{P}$ implies the presence of time-chiral phonons. Anything moving in time-chiral systems can induce electric polarization along the motion direction, thus acquiring time chirality. One intriguing characteristic of this time-chiral phonons is that phonons moving in the direction of the applied $E$ behave differently from those moving against it, exhibiting distinct behaviors or energy levels depending on the moving direction. The term "kinetomagnetism of chirality" [37] was first introduced to describe how motion in chiral systems can induce magnetization along the direction of motion, imparting chirality to moving objects, such as chiral phonons. The concept extends to time-chiral phonons, where the phenomena are referred to as "kinetoelectricity of time chirality". This exciting concept of time-chiral phonons will undoubtedly initiate theoretical and experiment explorations of time-chiral phonons. A particularly fascinating aspect is that this idea could be valid for any motion of quasi-particles, such as phonons and magnons, in time-chiral systems.

Second, $\mathcal{D}' \bullet \mathcal{A} = \mathcal{A}'$ indicate that shear stress, which behaves like ferro-rotation $\mathcal{A}$, can induce net magnetic moment $\mathcal{A}'$ in any time-chiral materials. In other words, they do exhibit shear-stress piezomagnetism (Fig. 2(b)). The shear force, depicted with green arrows, is associated with their rotational arrangement with two-fold symmetry around the out-of-plane axis, with all mirrors broken parallel to this axis, i.e., ferro-rotation axis. We also expect a reverse effect as shown in Fig. 2(e), i.e., applying magnetic field ($H$) can induce shear strain. One interesting experiment can be: applying $H$ on orthorhombic systems with time chirality such as *mmm*. In this case, the application of $H$ on *mmm* along $z$ may induce a uniform strain along the *xy* or *yx* direction, which can be related with shear strain along $z$.

Finally, $\mathcal{D}' \bullet \mathcal{C} = \mathcal{C}'$ indicates that when torsional stress (acting as $\mathcal{C}$)[38], as depicted as oppositely rotating green arrows on the top (solid) and bottom (dashed), is applied to any time-chiral states, the torsional stress can induce $\mathcal{C}'$ with longitudinal linear magnetoelectricity. Specifically, applying $E$ can induce net magnetic moments along the direction of $E$, and vice versa (Fig. 2(c)). Its reverse effect can be a unique behavior like this: when $E$ and $H$ are simultaneously applied to any time-chiral states, specimen twisting as chirality, can be induced (Fig. 2(f)). Thus, natural optical activity, in addition to the Faraday effect due to applied $H$, can be observed. Since natural optical activity is reciprocal, but the Faraday effect is nonreciprocal, their effects can be separated by measuring them with the flipped light propagation direction[39].

We have discussed three dot products related to time chirality based on symmetry, and now, we turn our attention to time-chiral domains, which should exhibit opposite behaviors. For instance, as shown in Fig. 2(f), time-chiral domains experience opposing torsional strains. Similarly, shear stress applied to a time-chiral state induces magnetization, whereas its **T**-state produces magnetization in the opposite direction. These phenomena in time-chiral states occur along any direction, with the relevant direction usually determined by the external vectorial stimuli, such as $E$, $H$, and shear stress. These effects can be harnessed to manipulate or control time-chiral domains by combining "external perturbations discussed above" with "the specimen's responses (i.e., observables). This is particularly relevant during cooling across time-chiral phase transitions. For example, applying both shear stress and a magnetic field along the same direction to a time-chiral system during cooling through the time-chiral phase transition temperature can induce a single, stable time-chiral domain.

### C. Time chirality in crystalline materials

In this section, we begin by examining time-achiral states with four head-to-tail magnetic toroidal spin configurations, as shown in Figs. 3(a) and 3(c), and magnetic monopoles in Fig. 3(b) and 3(d), all with tetramerized square unit cells outlined by black dashed squares. Although these configurations break **T**-symmetry, **T**-states can be restored in the free spatial operation condition as illustrated in Figs. 3(a-d). In contrast, orthorhombically distorted magnetic toroidal state (Figs. 3(e) and 3(f)) and orthorhombically distorted magnetic monopole state (Figs. 3(g) and 3(h)) can break the $C_{2xy}$ and $m_{xy}$ from the states in Fig. 3(c-d), and thus, form time-chiral states.

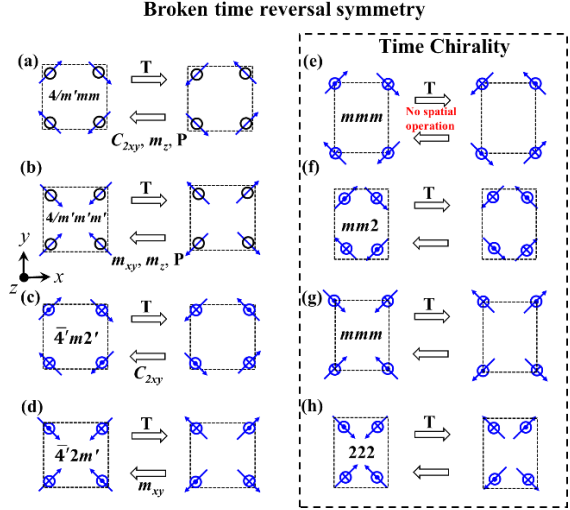

FIG 3. Exemplary broken **T**-symmetry magnetic states with Time Chirality and Time Achirality. Black dashed boxes outline unit cells or repeating units. (a) Toroidal state generated by four head-to-tail configuration of spins (left) with its **T**-state (right). The two states can be transformed into each other through spatial rotation ($C_{2xy}$), mirror reflection ($m_z$) or space inversion (**P**). The tetramerized configuration corresponds to MPG **4/m′mm**. (b) Magnetic monopolar state with a tetramerized unit cell of MPG **4/m′m′m′**, along with its **T**-state. The two states can covert through mirror reflection or space inversion. (c) Magnetic toroidal moment with alternating canted moments, corresponding to MPG $\bar{4}'m2'$ and its **T**-state (d) Magnetic monopole with alternating canted moments, corresponding to MPG $\bar{4}'2m'$ and its **T**-state. (a-d) time-achiral states. (e-f) Orthorhombically-distorted magnetic toroidal states from (c). (e) A non-tetramerized configuration corresponds to MPG **mmm**. (f) A tetramerized spin structure with an orthorhombic unit cell corresponds to MPG **mm2**. (g-h) Orthorhombically-distorted magnetic monopole states from (d). (g) A non-tetramerized spin structure corresponds to MPG **mmm**. (h) A tetramerized spin structure with an orthorhombic unit cell corresponds to MPG **222**. (e-h) time-chiral states.

The orthorhombically distorted spin states correspond to MPGs **mmm** (non-tetramerized, Fig. 3(e)) and **mm2** (tetramerized, Fig. 3(f)). The MPG **mm2** state has mirror reflections ($m_x$ and $m_y$) and spatial rotation ($C_{2z}$). The **P**, **T** and **PT** symmetries are broken when the free rotation condition is not considered. However, under the free rotation condition, the **mm2** effectively revert to **mmm** with unbroken **P** symmetry, while preserving a time-chiral nature. Both MPGs **mm2** and **mmm** fall under the time-chiral {**T**,**PT**} pink circle in the trinity diagram in Fig. 4. Another example is MPG **222** of Fig. 3(h), with $C_{2x}$, $C_{2y}$, and $C_{2z}$ symmetries, remains unchanged even under the free rotation condition. MPG **222** breaks {**P**,**T**,**PT**} symmetries. Examples of symmetry consideration, both with and without the free rotation condition are tabled in the Supplemental Materials S2

Using the method described, we identify 32 MPGs, so-called colorless MPGs or type-I MPGs, can belong to the time-chiral group, highlighted in Fig. 4. In the time chirality circle, centrosymmetry or non-centrosymmetry do not affect its classification. Time chirality has broken {**T**,**PT**} under the free rotation condition. Therefore, time chirality has broken {**T**,**PT**} under the free spatial operation condition, since any mirror operation is same with **P** under the free rotation condition along any directions and the **P** symmetry does not matter for time chirality as long as **PT** is broken.

Super-chirality group is defined to have broken {**P**,**T**,**PT**} under the free rotation condition, so **222** belong to the super-chirality group, which will be discussed in the next section. In addition to 11 super-chiral MPGs at the center, 5 MPGs out of the remaining 21 MPGs, are compatible with ferromagnetism ($\bar{1}$, **2/m**, **4/m**, $\bar{3}$, and **6/m**). The other 16 MPGs are associated with antiferromagnetism. For example, all states in Figs. 3(e-h) are both time-chiral spin structures and antiferromagnetic spin structures. Antiferromagnets, lacking any net magnetic moment, are insensitive to external magnetic fields, making them challenging to control. However, the time-chirality phenomena presented in Fig. 2 offer promising strategies for manipulating time-chiral and antiferromagnetic materials. Among these 16 antiferromagnetic MPGs, 10 MPGs (**m**, **mm2**, $\bar{4}$, **4mm**, $\bar{4}2m$, **3m**, $\bar{6}$, **6mm**, $\bar{6}m2$, and $\bar{4}3m$) exhibit piezoelectric properties and 6 MPGs (**mmm**, **4/mmm**, $\bar{3}m$, **6/mmm**, $m\bar{3}$, and $m\bar{3}m$) do not belong to either the piezoelectric or other categories.

Subgroup relationships among MPGs in Fig. 3 are highly informative. **mm2** [time chiral, Fig. 3(f)] is a subgroup of $\bar{4}'m2'$ [Fig. 3(c)], which is a subgroup of **4/m′mm** [Fig. 3(a)], and **222** [super-chiral, Fig. 3(h)] is a subgroup of $\bar{4}'2m'$ [chiral prime, Fig. 3(d)], which is a subgroup of **4/m′m′m** [Fig. 3(b)]. In addition, both **mm2** (time chiral) and **222** (super-chiral) are a subgroup of **mmm** (time chiral). In other words, time-chiral states tend to have lower symmetry, compared to broken **T**-symmetry spin structures that are non-time-chiral, in general.

**D. Time chirality and conjugate properties**

Following the discussion in section C, the definitions of the chiral and time-chiral states in 3D space become more succinct if we consider space inversion (**P**), time reversal (**T**) and **PT** symmetry. The trinity diagram shown in Fig. 4 also includes MPGs of chirality and chirality prime [40, 41] under the free rotation condition. In the free rotation condition, Chirality lacks {**P**, **PT**} symmetries, Time chirality has broken {**T**,**PT**} symmetries, and Chirality Prime has been defined as broken {**P**,**T**}[18, 30]. These criteria are clearly demonstrated in low-symmetry MPGs such as 1, 1′, $\bar{1}$, $\bar{1}'$, which represent four groups in the trinity diagram (Fig. 4). MPG 1 is super-chirality with broken {**P**,**T**,**PT**}. MPG $\bar{1}$ is linked to time chirality, showing broken {**T**,**PT**} and unbroken **P**. MPG 1′ relates to the circle of chirality with broken {**P**,**PT**} and unbroken **T**, and MPG $\bar{1}'$ belongs to chirality prime with broken {**P**,**T**} and unbroken **PT**. No MPGs fall into regions with only two out of three. 11 MPGs have all characteristics of chirality, time chirality and chirality prime. These are referred to super-chirality with broken {**P**,**T**,**PT**} under the free rotation condition, positioned at the center of the trinity diagram. Super-chirality is supposed to exhibit all properties and phenomena that are associated with chirality, time chirality and chirality prime.

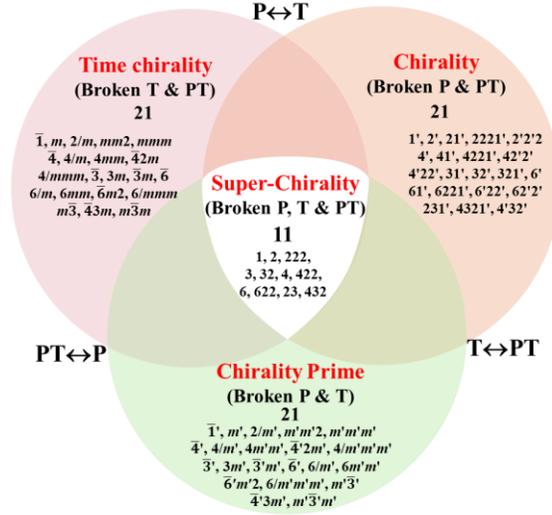

FIG 4. Trinity diagram of MPGs for chirality, time chirality, and chirality prime. The overlapping region of all of chirality, time chirality, and chirality prime is for super-chirality. Note that there cannot be any MPGs for only two out of three main circles. Their conjugate relationships with the exchange of **P**↔**T**, **T**↔**PT**, and **PT**↔**P** are shown.

Each circle is associated with its own characteristic physical properties and phenomena. For example, chirality is associated with various chiral activities such as natural optical activity, magnetochiral effect[19, 20, 42], and chiral phonons[23] with distinct left and right chiral features. Chirality prime exhibits longitudinal linear magneto-electric effects[18]. The MPG of the magnetic state of $Cr_2O_3$ [43] with diagonal linear magnetoelectricity is $\bar{3}'m'$, one of the chirality prime MPGs in Fig. 4. Furthermore, time chirality is associated with temporal nonreciprocity, manifesting in distinct behaviors for forward and backward time directions, which will be discuss in Section E.

Three circles (Fig. 4) have conjugate relationships in the following manner: chirality and time chirality can be switched to each other with **P**↔**T** swap, time chirality and chirality prime can be exchanged to each other with **PT**↔**P** switch, so does chirality prime and chirality with **T**↔**PT** exchange. Symmetry-driven conjugate properties have been observed in vector-like properties of magnetic, electric, and toroidal moments, along with their relevant properties[41, 44]. On the other hand, Fig. 4 demonstrates the scalar-type trinity diagram.

We emphasize the orientational independence of properties associated with chirality, time chirality, and chirality prime, meaning that the directions of measurable responses are determined by the direction of external stimuli rather than the intrinsic orientation. While chirality is often associated with twisted structures or screw axes in crystallography, it generally lacks a specific direction. The relevant direction for chiral activities in systems is determined by external perturbations, such as light propagation for natural optical activity or the applied ***H*** in the magnetochiral effect. Similarly, in chiral prime systems, diagonal linear magnetoelectricity occurs in any direction, with induced magnetization (or polarization) aligned to the applied ***E***/***H*** field.

**E. Temporal nonreciprocity**

In this section, we focus on temporal nonreciprocity[10], which arise from the genuine breaking of **T**-symmetry under the free rotation condition. Temporal nonreciprocity is defined as a situation where the experimental setup and observables are not mutually reciprocal over time, revealing a time-dependent coupling that breaks the temporal reciprocity. Our consideration focuses only on (quasi-)equilibrium states with no obvious net fluxes. Fig. 5 provides a practical explanation of how temporal reciprocity works or not. First, Figs. 5(a-c) show the NDD effect [45] that arises from magnetochiral effect in a chiral materials when subjected to an external magnetic field (***H***). On the left side of '≈' symbol, the time (*t*)-evolving experimental setup (+***H*** with "linearly" increasing H magnitude with time in the blue box) is shown, while the right side illustrates the *t*-evolving observables, i.e. increasing NDD (***k***). $VNb_3S_6$ is one of the chiral magnets[46] with MPG **2′2′2**. It has symmetries $C_{2z}$, $C_{2x}T$ and $C_{2y}T$ with broken **T**-symmetry. This results in the formation of two types of ferromagnetic domains colored orange and light blue (Figs. 5(a) and 5(b)). These two domains are time-achiral, meaning they can overlap via $C_{2x}$ or $C_{2y}$ operations.

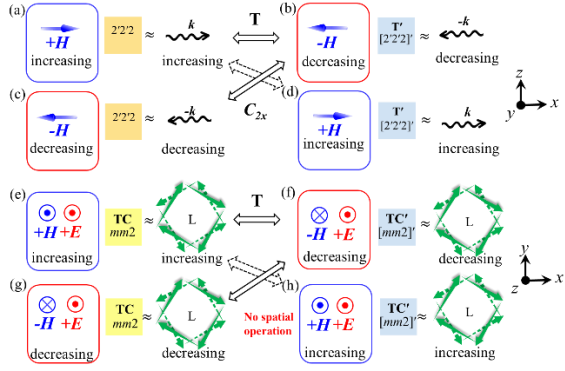

FIG. 5. Temporal (non)reciprocity (a) *t*-evolving Nonreciprocal Directional Dichroism (NDD) in chiral MPG **2′2′2**, showing a *t*-evolving experimental setup (blue box) with an external magnetic field (***H***), the initial state (orange), and the corresponding observables with light (***k***) along the *x*-axis. (a) and (b) are linked though **T**-symmetry: (b) has the **T**-reversed experimental setup (red box) and the **T**-state (T′, light blue) and the corresponding **T**-reversed observable. (c) and (b) relates through the $C_{2x}$, showing that the **T**-state (T′, light blue) returns to its original state (orange) in the presence of the **T**-reversed experimental setup and the **T**-reversed observables, thus the experiment exhibits temporal reciprocity. (d) and (a) are linked through $C_{2x}$. The terms "increasing" and "decreasing" refer only to the magnitude, not the sign. (e-h) illustrate the left(L)-handed twisting responses in time-chiral MPG ***mm2*** under external ***E*** and ***H***. (e) and (f) are linked though **T**-symmetry: (e) has a *t*-evolving experimental setup (blue) with external +***E*** and +***H***, the initial time-chiral (TC, yellow) state, and the resulting twisting, and (f) has the **T**-reversed *t*-evolving experimental setup (red), the **T**-reversed time-chiral (TC′, light blue) state, and the **T**-reversed twisting. Temporal nonreciprocity occurs because no spatial operation links (f) to (g), nor (e) to (h).

The relation between Fig. 5(a) and 5(b), or Fig. 5(c) and Fig. 5(d) can be viewed as time-reversed (**T**-reversed) processes, like video-playing-forward or -backward. The terms of 'increasing' and 'decreasing' marked in Fig. 5 refer the linear changes of the magnitude, without considering the sign. **T**-reversed Fig. 5(a) leads to Fig. 5(b) with "**T**-reversed *t*-evolving experimental setup" (decreasing -***H*** in the red box), "**T**-reversed observables" (decreasing -***k***), and "**T**-state" (light blue state). However, $C_{2x}$ on Fig. 5(b) leads to Fig. 5(c) with "**T**-reversed *t*-evolving experimental setup", "**T**-reversed observables", and initial state (orange state). Thus, the process on an identical specimen is **T**-symmetric, and the system is temporal reciprocal, consistent with the unbroken **T**-symmetry in chiral MPG **2′2′2** under the free rotation condition (i.e., it is not time-chiral). Fig. 5(a) and Fig. 5(d) represent a single time (*t*)-evolving experimental setup applied to a different domain.

In contrast, time-chiral systems become **T**-asymmetric (i.e., temporal nonreciprocal), as shown in Figs. 5(e) and 5(g). As illustrated in Fig. 2(f), applied ***E*** and ***H*** induce torsional strain in opposite time-chiral domains (TC yellow, and TC′ light blue). Fig. 5(e), a *t*-evolving stimulus (increasing +***E*** +***H*** in the blue box) on the yellow TC domain induces Left-handed twisting signal as the *t*-evolving observable. However, when applied a **T**-reversed *t*-evolving stimulus (decreasing +***E*** -***H*** in the red box of Fig. 5(f)) and **T**-reversed *t*-evolving observable (decreasing Left-handedness twisting of Fig. 5(f)), no spatial operation can link TC′ (light blue state of Fig. 5(f)) to its counterpart TC (yellow state of Fig. 5(e)). This lacks any symmetry connection between Figs. 5(f) and 5(g) or Fig. 5(e) and Fig. 5(h) aligns with the symmetry criteria of time-chiral states, where only **T**-symmetry and no other spatial or combined operations can map TC and its counterpart TC′, which leads to temporal nonreciprocity.

In other words, first imagine taking a video of the experiment with "*t*-evolving experimental setup" on "a time-chiral state", and "*t*-evolving observables". Now, if we perform the next experiment with the video-playing-backward version of "the *t*-evolving experimental setup" on the same time-chiral state, then the result will be different from the video-playing-backward version of the initial result of "the *t*-evolving observables" – this should work along any spatial orientations or even in mirror images. If we perform the initial *t*-evolving experiment now on the **T**-reversed time-chiral state, then the result will be also different from the original result of "the *t*-evolving observables", since the **T**-reversed time-chiral state cannot be linked to the initial time-chiral state without invoking **T**-symmetry. Further details on the symmetry of temporal nonreciprocity are provided in the Supplemental Materials Fig. S1.

The exact nature of Temporal nonreciprocity on time chiral states remains to be explored. It may manifest as the asymmetric *t*-evolving twisting strength or speed. It could also be linked to the asymmetric nucleation and growth of opposite time-chiral domains, which can be reflected, for example, in the distinct slopes in the Kibble-Zurek scaling. This scaling describes the time evolution of topological defects generated when driving a system through a critical point[47-50].

Note that sometimes, "temporal nonreciprocity" refers to an irreversible change of certain materials'

properties to external stimuli, highlighting, for example, how their interaction with light evolves non-reciprocally over time. However, we focus on temporal nonreciprocity in equilibrium states or quasi-equilibrium state in the absence of any loss/gain.

### F. Candidate materials for time chirality

There exist numerous magnetic materials where the time-chiral physical properties and phenomena that we have discussed can be explored. For example, the *mmm* state has been reported in MnTe ($T_n$=300 K)[51], LaCrO$_3$ ($T_n$=290 K)[52], and Fe$_{1.5}$Mn$_{1.5}$BO$_4$ ($T_n$=100 K)[53]. Below 125 K, FePO$_4$ exhibits the time-chiral and chiral transition with the MPG **222** [54], while FeSb$_2$O$_4$ has been reported to form ***mm2*** state with $T_n$=45 K [32, 55]. Multiferroic Ca$_3$Co$_{2-x}$Mn$_x$O$_6$ [56] has a ferroelectric and time-chiral transition with the MPG **3*m*** with $T_n$=16 K. Chiral skyrmion Cu$_2$OSeO$_3$ [57] exhibits a polar and time-chiral transition at $T_n$=60 K with MPG **3**. Many other compounds[32] also feature relevant MPGs for both pure time chirality as well as super-chirality. Finally, note that since various broken symmetries are involved in chirality, time chirality, chirality prime and super-chirality, the relevant various domains, their domain topology, and the mutual coupling among different types of domains will be an important topic to be investigated.

## IV. FUTURE PROSPECTS

The concept of time chirality and temporal nonreciprocity, introduced in this paper, opens up exciting possibilities for future research in both fundamental and applied physics. As we have demonstrated, time chirality, distinct from the conventional broken **T**-symmetry, brings with it the potential for uncovering a range of exotic phenomena, such as **E**-induced NDD, shear-strain piezomagnetism, induction of torsional strain in the presence of simultaneous **E** and **H**, and time-chiral phonons, most of which were previously overlooked in traditional **T**-symmetry breaking frameworks. The key foundational principle that underpins these phenomena is the orientational independence of broken {**T**, **PT**} symmetries. Specifically, where **T**-symmetry is broken, it must remain broken even under the operation of any spatial symmetry operations, mirror reflections, space inversion, translation and spatial rotations. We can refer to this as true **T**-symmetry breaking, drawing an analogy with the distinction between true and false chirality described in literature [14].

False chirality, as defined by Barron[14], refers to systems that appear chiral under parity (**P**) but are not truly chiral because their mirror-image states can be connected through the combined operation **PT**. Thus, breaking **P** and **T** symmetries, while preserving **PT** or certain spatial combinations, characterizes a group that we denote as chirality prime (illustrated in Fig. 4) and corresponds to Barron's definition of false chirality. In contrast, true chirality requires the absence of both **P** and **PT** symmetries. Time chirality emerges as the complementary case in this symmetry framework, completing the set when considering the conjugate relationships among true chirality, chirality prime (false chirality), and time chirality. We emphasize that chirality prime has its own characteristic physical property of "linear magnetoelectricity", so the term of "false chirality" may not be appropriate.

Most of these phenomena have been scarcely addressed, both theoretically and experimentally. Temporal nonreciprocity in time-chiral states describing the time-asymmetric behavior is another concept that will open a fresh avenue for the true nature of time chirality. For example, for time-chiral phonons, those propagating in the direction of the applied electric field exhibit distinct behavior compared to those moving in the opposite direction, along with other phenomena previously discussed. An especially intriguing aspect of this concept is its potential applicability to the motion of any quasi-particles, such as phonons, magnons, and electrons, in time-chiral systems. The evolution of time-chiral domains under time-varying external perturbations presents an intriguing concept, probably distinct from the notion of opposite time flows in the universe and anti-universe. However, various theoretical and experimental aspects, such as the influence of different cooling rates across time-chiral transition temperatures on time-chiral domain evolution, warrant further investigation.

Time-chiral states and temporal nonreciprocity, can occur in artificial metamaterials, complex molecule systems, and crystalline magnetic materials. In practice, our identification and classification of MPGs for time chirality will be the first and most important guidance to explore time-chiral materials. Particularly, many MPGs are associated with antiferromagnets, which are insensitive to external magnetic fields, making them challenging to manipulate. However, the time-chirality phenomena highlighted in our work (Fig. 2) promise strategies for observing and controlling time-chiral and antiferromagnetic materials[58].

While the apparent coupling between magnetic order, light, shear strain, and the coexistence of magnetoelectric fields in time-chiral materials might seem straightforward from the symmetry analysis perspective, the potential for discovering novel and unforeseen effects invites significant theoretical and experimental exploration. As research into time chirality and temporal nonreciprocity continues, the scope for innovative applications becomes increasingly vast for both fundamental science and practical technological development.

## ACKNOWLEDGMENTS

We greatly appreciate the valuable discussions we've had with Kai Du, Seong Joon Lim, and David Vanderbilt. SWC acknowledges the Global Fellowship from the University of St. Andrews, which supported this work during SWC's visit to the university in 2024. This work was supported by the W. M. Keck foundation grant to the Keck Center for Quantum Magnetism at Rutgers University.

------------------------------------------------